\documentclass{emulateapj}
\usepackage{apjfonts}

\shorttitle{Evolution of compact galaxies}
\shortauthors{Poggianti et al.}

\begin{document}


\title{The evolution of the number density of compact galaxies}


\author{B.M. Poggianti$^1$, A. Moretti$^2$, R. Calvi$^{1,2}$, M. D'Onofrio$^2$, T. Valentinuzzi$^2$, J. Fritz$^3$, A. Renzini$^1$}
\affil{$^1$INAF-Astronomical Observatory of Padova, Italy,
$^2$Astronomical Department, University of Padova, Italy,
$^3$Sterrenkundig Observatorium Vakgroep Fysica en Sterrenkunde Universiteit Gent, Belgium
}



\begin{abstract}
We compare the number density of compact (small size) massive galaxies
at low and high redshift using our Padova Millennium Galaxy and Group
Catalogue (PM2GC) at $z=0.03-0.11$ and the CANDELS results from Barro
et al. (2013) at $z=1-2$.  The number density of local compact
galaxies with luminosity weighted (LW) ages compatible with being
already passive at high redshift is compared with the density of
compact passive galaxies observed at high-z.  Our results place an
upper limit of a factor $\sim 2$ to the evolution of the number
density and are inconsistent with a significant size evolution for
most of the compact galaxies observed at high-z.  The evolution may be
instead significant (up to a factor 5) for the most extreme,
ultracompact galaxies.  Considering {\it all} compact galaxies,
regardless of LW age and star formation activity, a minority of local
compact galaxies ($\leq 1/3$) might have formed at $z<1$.  Finally, we
show that the secular decrease of the galaxy stellar mass due to
simple stellar evolution may in some cases be a non-negligible factor
in the context of the evolution of the mass-size relation, and we
caution that passive evolution in mass should be taken into account
when comparing samples at different redshifts.
\end{abstract}





\section{Introduction}

The average size of passive and massive galaxies at $z=1-2.5$ has been
observed to be much smaller than that of galaxies of similar masses in
the local Universe (Daddi et al. 2005, Trujillo et al. 2006, Cimatti
et al. 2008, van Dokkum et al. 2008, Cassata et al. 2011, Damjanov et
al. 2011, to name a few), usually adopting as comparison low-z
galaxies with a Sersic index greater than 2.5 (Shen et al. 2003).
These results have suggested that galaxies have undergone a strong
evolution in size.  Minor dry mergers have been proposed as the main
mechanism driving such evolution (Naab et al. 2009, Oser et al. 2012).

At low redshift, compact massive galaxies are a small fraction of the
overall population in the general field (Trujillo et al. 2009, Taylor
et al. 2010, Poggianti et al. 2013, hereafter P13) but are much more
common in galaxy clusters (Valentinuzzi et al. 2010, hereafter V10).

In P13, we have studied the population of compact massive
galaxies in the local universe for the first time on a non-Sloan
sample representative of the general field galaxy population, 
the Padova Millennium Galaxy and Group catalogue
(PM2GC, Calvi et al. 2011).  
Compact PM2GC galaxies with radii and mass densities
comparable to high-z massive and passive galaxies are mostly S0s
with intermediate-to-old stellar populations, and have the
characteristics to be likely little-evolved descendants of high-z
compact galaxies (P13).

At a given mass, galaxies with older stellar populations are on
average more compact (van der Wel et al. 2009, Saracco et al. 2009, 
V10, P13 and references
therein). When high-z passive galaxies are compared with local
galaxies {\it with old stellar populations}\footnote{Luminosity-weighted
ages compatible with being already passive at high-z.} (their most probable
descendants), the average evolution in size at fixed mass
between  $z=1-2.5$
and $z\sim 0$ turns out to be mild, a factor $\sim
1.6$, as half of the evolution observed for the overall passive galaxy
population is driven by larger galaxies becoming passive at lower
redshifts (P13, V10).

A crucial aspect for assessing the real evolution in size of
individual galaxies is to evaluate how many of the distant galaxies
have remained compact till today, comparing high- and low-z number
densities.  Cassata et al. (2011) and (2013) have studied the evolution of the
number density of passive early-type galaxies and concluded this is
driven by both the size growth of compact galaxies and, mostly,
by the appearance
of new early type galaxies with larger sizes. 
Similar conclusions are reached by Carollo et al. (2013), who
find no change in the number density of compact quenched early-type
galaxies of masses $<10^{11} M_{\odot}$, and a 30\% decrease
at higher masses over the redshift interval $0.2<z<1$.
Saracco et al. (2010),
comparing their high-z morphologically-selected 
early-type sample with the cluster galaxies by
V10, concluded that most of the compact early-types at high-z can be
accounted for by their local counterparts.
In contrast, Taylor et al. (2010) found a strong evolution
in the number density of extremely compact galaxies using
as local comparison disk-free red sequence SDSS galaxies.

Recently, new estimates for the number density of compact high-z
galaxies have become available for the GOODS-S and UDS fields of
CANDELS (Barro et al. 2013, hereafter B13).  This sample is
unprecedented for number of galaxies and quality, and 
relies on selection criteria that we can try to
reproduce at low redshift.
In this paper we compare the CANDELS results with our PM2GC
local sample to establish the amount of evolution in the number density
of compact galaxies. This directly constrains the number of galaxies
that experienced a significant evolution in size.

We use ($H_0$, ${\Omega}_m$,
${\Omega}_{\lambda}$) = (70,0.3,0.7), and a Chabrier (2003) IMF, to
be consistent with the high-z dataset.

\section{Data and methodology}

Our aim is to compare our low redshift data
with the high-z results from B13, who studied
the structure of massive galaxies at $z=1.4-3$ in the GOODS-S and UDS
CANDELS fields. We use the number densities obtained by B13 
for all compact galaxies and for
passive compact galaxies 
with masses above $10^{10} M_{\odot}$, for which their sample is 90\%
complete. B13 define passive (``quiescent'') galaxies
to have a specific star formation
rate $<10^{-0.5} \rm Gyr^{-1}$, corresponding to a star formation rate
about 1/10 of that of typical star-forming 
galaxies at $z=2$ on the star formation rate-mass relation.
Their circularized galaxy radii were obtained with GALFIT
from $HST/WFC3$ $H$-band\footnote{This corresponds to rest-frame
wavelengths from B to R band. Galaxy sizes are larger at shorter
wavelengths (P13), therefore by using local B-band sizes we obtain a
conservative upper limit to the density evolution.} images (van der
Wel et al. 2012), and their masses with a spectrophotometric model
based on Bruzual \& Charlot (2003).

Our low-z analysis is based on the PM2GC, a spectroscopically complete
sample of galaxies at $0.03 \leq z \leq 0.11$ brighter than $M_B <
-18.7$. This sample is sourced from the Millennium Galaxy Catalogue
(MGC, Liske et al. 2003, Driver et al. 2005), a B-band contiguous
equatorial survey complemented by a 96\%
spectroscopically complete survey down to B=20.  The PM2GC is similarly
complete for masses $M_{\star} \geq 1.6 \times 10^{10} M_{\odot}$, and contains
1515 galaxies above this limit.  

The image quality and the
spectroscopic completeness of the PM2GC
are superior to Sloan, and these qualities
make it an interesting dataset to study galaxy sizes in a complete
sample of galaxies at low-z, as outlined in P13. Effective-radii,
axial ratios and Sersic indexes were measured on MGC B-band images
with GASPHOT (Pignatelli \& Fasano et al. 2006, Bindoni et al. in
prep.), an automated tool which performs a simultaneous fit of the
major and minor axis light growth curves with a 2D flattened
Sersic-law, convolved by the appropriate, space-varying PSF. Details
on the size measurements and a comparison with independent size
estimates of PM2GC galaxies are given in P13. Galaxy stellar mass
estimates were derived by Calvi et al. (2011) using the Bell \& de
Jong (2001) relation and are in good agreement with
DR7\footnote{http://www.mpa-garching.mpg.de/SDSS/DR7/Data/stellarmass.html}
(Abazajian et al. 2009)
masses, with no offset and a $<0.1$dex scatter (P13).

\subsection{Sample selection: compactness criterion}

We stress that no morphological information is used in the galaxy
selection, neither at high- nor at low-z.

B13 define compact those galaxies with 

\begin{equation}
\rm
log(M_{high-z}/r_e^{1.5}) \geq 10.3 \, M_{\odot} kpc^{-1.5}
\end{equation}

where
$M_{high-z}$ is the stellar mass of high-z galaxies and $r_e$ their
effective radius.

The stellar mass of a system is defined as the sum of the mass in
living stars + the mass of stellar remnants\footnote{The stellar mass
in this paper and in general in the literature is described by
equation (2) in Longhetti \& Saracco (2009).}.  Assuming passive
evolution, the stellar mass of a galaxy changes with time simply due
to the evolution of its stars: as they progressively evolve and
eventually die, they retain only part of their mass as remnant.


For a stellar generation of solar metallicity, a Chabrier IMF and the
Bruzual \& Charlot (2003) model,  the fraction of
initial stellar mass that remains is equal to 1 for ages less than $1.9 \times
10^6$ yr, while it can be approximated as
 $f(t) = 1.749 - 0.124 * log \, t$ at older ages, where $t$
is the age of the stellar population in years.

If a high-z galaxy is observed very soon after the end of the bulk of
star formation, say $10^8$ yr after, by $z \sim 0.1$ it retains only $\sim
70\%$ of its observed high-z mass.
If star formation stopped 2 Gyr before the high-z observations (at
z=4.8 for a z=2 observed galaxy), this fraction is $\sim$90\%. We adopt an
intermediate value of 80\%, corresponding to a quenching of star
formation occurred 0.6 Gyr before the high-z observations (at z=2.5
for a z=2 observed galaxy), in line with recent observations of
recently quenched galaxies at $z=1-2$ (Wuyts et al. 2010, Whitaker et
al. 2012). Adapting eqn.~1, out compactness threshold at
low redshift is

\begin{equation}
\rm
log(M_{low-z}/r_e^{1.5}) = \rm log(0.8 \times M_{high-z}/r_e^{1.5}) =
10.2 \, M_{\odot} kpc^{-1.5}
\end{equation}

It is worth noting that passive evolution in mass 
will also lead to larger sizes for adiabatic expansion
(Hills 1980) if the mass returned by stars to the interstellar medium
is lost by galaxies in a SN- and/or AGN-driven galactic wind,
as indicated by X-ray data for early-type galaxies
(Ciotti et al. 1991). Such mechanism is internally driven and
may contribute to the evolution in size. The effects on galaxy sizes of
rapid mass loss driven by quasar superwinds in massive galaxies
and by supernova-driven winds in less massive galaxies
have been discussed by Fan et al. 2008. 
One type of galaxies where there is unequivocal evidence
for merger-driven evolution both in mass and in size
are the Brightest Cluster Galaxies (eg. Lidman et al. 2012,
Valentinuzzi et al. 2010b, but see also Stott et al. 2010).
However, such galaxies may have a very different evolution from the
rest of the massive galaxy population (P13).

Finally, if B13 used Maraston (2005) model instead of Bruzual \& Charlot
(2003), their high-z masses would be typically $\sim$0.15 dex smaller
due to the different treatment of the TP-AGB stellar phase, which is
important for stellar population ages typical of high-z galaxies, but
is uninfluential on the masses at low-z.  As a consequence, adopting Maraston's
model, the compactness density threshold at low-z would shift to $\sim 10.05 \,
\rm M_{\odot} kpc^{-1.5}$.  In the following we will use the
compactness criterion based on Bruzual \& Charlot, unless otherwise
stated, but we will show that this is a major source of uncertainty in
the comparison of high- and low-z samples.
Above the PM2GC mass limit there are 141 compact galaxies according to
the compactness criterion based on Bruzual \& Charlot, and 294 using 
Maraston.


\subsection{Sample selection: LW ages}

All PM2GC galaxies have a spectrum from the SDSS, the 2dFGRS (Colless et 
al. 2001) or
the MGCz (Driver et al. 2005).  The galaxy stellar history was
derived by fitting the spectra with the spectro-photometric model
described in Fritz et al. (2007, 2011).  All the main
spectro-photometric features (continuum flux and shape,
and emission and absorption lines) are
reproduced by summing the theoretical spectra of Simple Stellar
Populations (SSPs) of 12 different ages,
from $3 \times 10^6$ to $\sim 14 \times 10^9$ years (P13).  
From the spectral analysis, it is possible to
derive an estimate of the average age of the stars in a galaxy
weighted by the light we observe. The LW
age was computed by weighing the age of each SSP composing the
integrated spectrum with its bolometric flux.
For passive galaxies, this
reflects to a first approximation the epoch of the last 
star formation episode.

The model age determination can be quite uncertain, especially at the old
ages considered here, and it is necessary to evaluate how these uncertainties
affect the number density estimates.

The spectrophotometric model obtains the best fit star formation history,
and consequent LW age, by performing a random exploration of the
parameter space, searching for the combination of star formation rate
and extinction values of the various single stellar populations
that minimizes the differences between the synthetic and the observed
spectra.

As explained in Fritz et al. (2007), the internal error on the LW age
is estimated by exploiting the characteristics of the optimization
routine. The path towards the best fit parameters depends on their
initial value, and so does the final solution (star formation history
and, consequently, LW age).  We perform 11 simulations for each
metallicity, and explore three metallicities, changing both the
initial parameters and the seed of the random number generator,
sampling the entire space of solutions.
The LW age error is taken to be half the difference between the highest
and lowest values of LW age among all simulations.
Such error is less than 1 Gyr for all compact PM2GC galaxies except for a 
few cases, with a median of 0.3 Gyr (see left panel of Fig.~1).

\begin{figure}
\centerline{\includegraphics[scale=0.48]{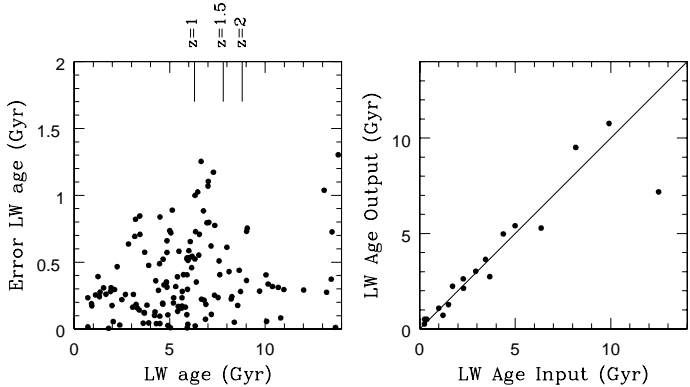}}
\caption{
{\bf Left.} Error on LW age for all compact PM2GC galaxies estimated as
explained in \S2.2.1.
{\bf Right.} Comparison between the LW age of simulated templates of 
different star formation histories (``input'') and the LW age recovered
for them from the model fitting (``output''). The templates include a 
wide range of star formation histories, from those mimicking the histories
of galaxies of different morphological types (ellipticals to late spirals),
to post-starbursts, to constant star formation rates truncated at different
times.
}
\end{figure}

To further test the age uncertainty, we used synthetic spectra
of galaxy templates with different star formation histories, 
mimicking the histories
of galaxies of different morphological types (ellipticals to late spirals),
post-starbursts and constant star formation rates truncated at different
times, as was done in Fritz et al. (2007).
The ``true'' (input) LW age of such templates are compared to the
LW age recovered by the spectral fitting
in the right panel of Fig.~1.
Generally, the model is able to recover the input LW age with a good
approximation, except for our oldest template for which it 
underestimates it. For our purposes, it is important that the model
does not show a tendency to {\sl overestimate} the LW ages, 
thereby assigning old ages to young galaxies.

In this paper we want to compare the number density of
compact passive galaxies at high-z
with that of equally compact low-z galaxies whose LW age testifies that they
were already passive at high-z, and therefore should comply with the
B13 criterion for passivity.  To do this, we select galaxies at $z\sim 0.1$
with a LW age equal to or greater than the time elapsed between three high
redshift values ($z=1,1.5$ and 2, corresponding to lookback times 6.3,
7.8 and 8.8 Gyr, respectively) and $z\sim0.1$. 

Visually inspecting the spectra of all 39 (59 for Maraston)
compact PM2GC galaxies with LW
ages $\geq 6.3$ Gyr (z=1), their passivity at the time they are
observed is confirmed by the lack of emission lines: only 5 (7 for Maraston)
galaxies
have weak (equivalent width [OII] $<5$ \AA) emission with line ratios 
consistent with a weak AGN.

Figure~2 shows that the stellar mass distribution of compact passive
 high-z galaxies is similar to the mass distribution of PM2GC compact
 and old (LW age corresponding to $z>1$) galaxies, and a KS-test is
 unable to prove there are significant differences (P=0.1). This is
 consistent with the hypothesis that PM2GC compact and old galaxies
 are local counterparts of the high-z compact and passive population.

\begin{figure}
\centerline{\includegraphics[scale=0.5]{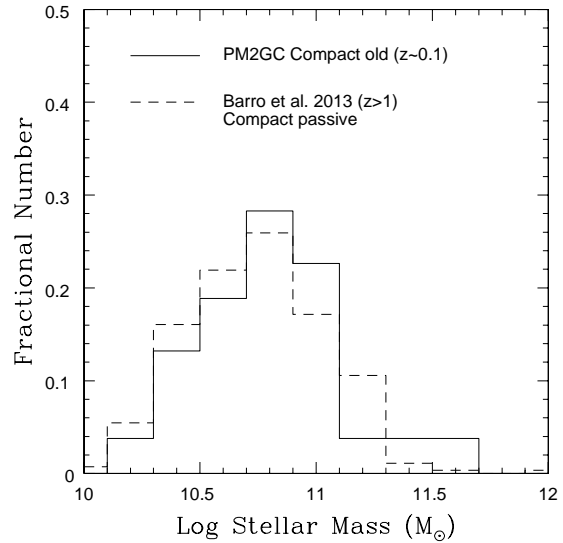}}
\caption{Stellar mass distributions of PM2GC compact and old galaxies
(solid histogram, LW ages $\geq$ than the lookback time to $z=1$) and 
of B13 high-z
compact passive galaxies (dashed histogram).
}
\end{figure}


Matching progenitors and descendants is one major problem
in trying to assess the number density evolution of any type
of high-z galaxies, including compact
galaxies. Different methods have been proposed to obtain a meaningful 
comparison, and none of them is free from problems. For example,
a morphological selection of early-type galaxies both at high-
and low-z does not take into account the fact that a large number
of high-z late-type galaxies turn into early-type galaxies by z=0, 
while using a fixed number density is prone to partial
contamination and relies on semi-analytic models
that do not reproduce the galaxy stellar mass function at redshifts above zero
(Leja et al. 2013).

Our approach based on LW ages is not free from problems either,
because it relies on the assumption that compact passive galaxies at
high-$z$ have not been "rejuvenated" by star formation at subsequent
times. Indeed, while it is reasonable to assume that all local compact
galaxies with old LW ages must have been compact and passive at
high-$z$, not all compact passive galaxies at high-$z$ necessarily
have old LW ages today. There might be some of their local
counterparts that are still compact, but have not remained always
passive throughout their evolution, as at later times they may have
experienced some episode of star formation such to disqualify them from
being "old" in the local sample. Though not optimal, our method links the
high-$z$ population with at least some of their low-$z$ counterparts,
thus providing an upper limit to the evolution of the number density
of compact passive galaxies, i.e., to the difference between the
high-$z$ and low-$z$ number of such galaxies.


\subsection{Number densities}

The high-z number densities for passive, star-forming and all galaxies
are taken directly from B13. They are for a mass limit $10^{10}
M_{\odot}$, which is slightly lower than the $1.6 \times 10^{10}
M_{\odot}$ PM2GC limit.  This difference should have little effect on
the number densities, as most of the high-z compact galaxies are well above
the PM2GC limit (Guillermo Barro 2013, private communication). The different
limit in mass will, again, result in underestimating the number density
of local compacts relative to that of high-z compacts.

For the PM2GC we compute the number density of compact galaxies of
different LW ages as described above, as well as the total number
density of all low-z compact galaxies, regardless of their age.  The
number of galaxies is divided by the comoving volume of the PM2GC
survey, that extends over an effective area of 30.88deg$^2$ on the sky
between $z=0.03$ and $z=0.11$. No volume correction is needed, as the
survey is complete down to our mass limit at $z=0.11$.
Our number densities for both Bruzual \& Charlot
and Maraston high-z models are given in Table~1.

\section{Results}

Figure~3 shows the mass-size diagram of all PM2GC galaxies above our
mass limit. The solid and dashed lines represent the compactness
threshold of B13 with and without the mass evolution effect, respectively.
The compactness criterion isolates galaxies approximately 2$\sigma$
away from the median PM2GC mass-size relation, as the plot shows.
This compactness criterion is slightly less strict than the limit for
superdense galaxies adopted in P13 (see green points in Fig.~3).

\begin{figure}
\centerline{\includegraphics[scale=0.48]{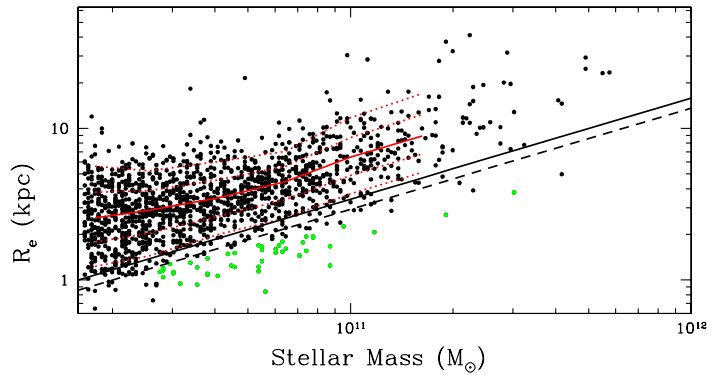}}
\caption{Circularized effective radius as a function of stellar mass
for all galaxies in the PM2GC mass-limited
sample (all points, 1515 galaxies). The solid and dashed black lines
represent the Barro's compactness criterion limit taking into account and
ignoring the effect of mass evolution, respectively (see text).
The red solid line, with dotted 1$\sigma$ and 2$\sigma$ lines, 
is the PM2GC median. Green points represent superdense galaxies
according to the compactness criterion adopted in P13.
}
\end{figure}

Our main result is presented in Fig.~4, where the number density of
low redshift compact galaxies is compared with the high redshift
values. Selecting only PM2GC compact galaxies with LW ages at least as old
as the lookback time between $z\sim 0.1$ and $z=1,1.5$ and 2 
we plot their number density values as red circles at the corresponding
redshifts in the left panel.

The largest uncertainty in the PM2GC number densities arises from the
precision of the LW age estimate obtained by the spectrophotometric
model. 
The number density error bars in Fig.~4 take into account the change in number
density varying the LW age corresponding to each redshift by $\pm 1$ Gyr.
As discussed in \S2.2.1, this corresponds to the upper envelope
of the internal age uncertainty, and we have no reason to 
believe there should be any systematic overestimation of ages
in the modeling. 
Such error dominates over the poissonian error.
The X-axis error bar shows the $\pm1$ Gyr intervals in redshift.

From Fig.~4, 
the number ratio of high- to low-z densities is $\sim 1$ at $z=1$ (no
evolution) and only $\sim 2.3$ at $z=2$. The low-z 
estimates are within 1-1.5 $\sigma$ of
the high-z estimates.

The red squares in the plot show the PM2GC number densities 
obtained adopting Maraston (2005) model at high-z, hence shifting the
high-z masses and consequent compactness threshold by 0.15 dex.  In
this case, high- and low-z number density values are even closer,
their ratio being = 1.6 at $z=2$.

We stress once again that, in deriving the number density at low-z,
only compact galaxies that have remained passive since high-z are
counted.  Some high-z compact quiescent galaxies might have remained
compact till today but have experienced star formation after they were
observed at high-z, and these are not considered in our low-z compact
galaxy census. Our number of old compact galaxies at low-z is
therefore potentially a lower limit to the number of compact quiescent
galaxies that have remained compact. Consequently, the difference
between the high-z and low-z values in Fig.~4 provides an upper limit
to the number density evolution.

Although an exact estimate of the evolution is hampered by the various
uncertainties involved, clearly a strong evolution in the
number density of compact galaxies, such as the factor 10 sometimes
quoted from simulations (Hopkins et al. 2009), 
is not supported by our analysis.
Our results are consistent with a small or even negligible evolution in the
number density of compact galaxies, suggesting that at most
$\sim$half 
of the high-z
compact galaxies have evolved in size by $z=0$.

At low-z, the total number density of compact galaxies in the PM2GC,
regardless of their LW age, is $4.7 \times 10^{-4} \, \rm Mpc^{-3}$.  
This value is compared in the right panel of Fig.~4 with the total number
density of compact passive + star-forming galaxies in CANDELS, which
reaches a maximum value $2.9 \times 10^{-4} \, \rm Mpc^{-3}$ at $z=1.6$.
This comparison suggests that the bulk of the compact galaxies
have formed their structure and their mass by $z\sim 1-1.5$, and that
an additional minority of them, up to about 1/3 of today's compact
galaxies, might have formed at lower redshifts.

This comparison also shows 
that the total number density of PM2GC compact galaxies
of any age is significantly higher than the number density of
passive compact galaxies at high-z. Relaxing the restrictive assumption
of no subsequent star formation, the whole population of high-z compact 
passive galaxies could evolve into the local compact population.

\begin{figure*}
\vspace{-10cm}
\centerline{\includegraphics[scale=1.0]{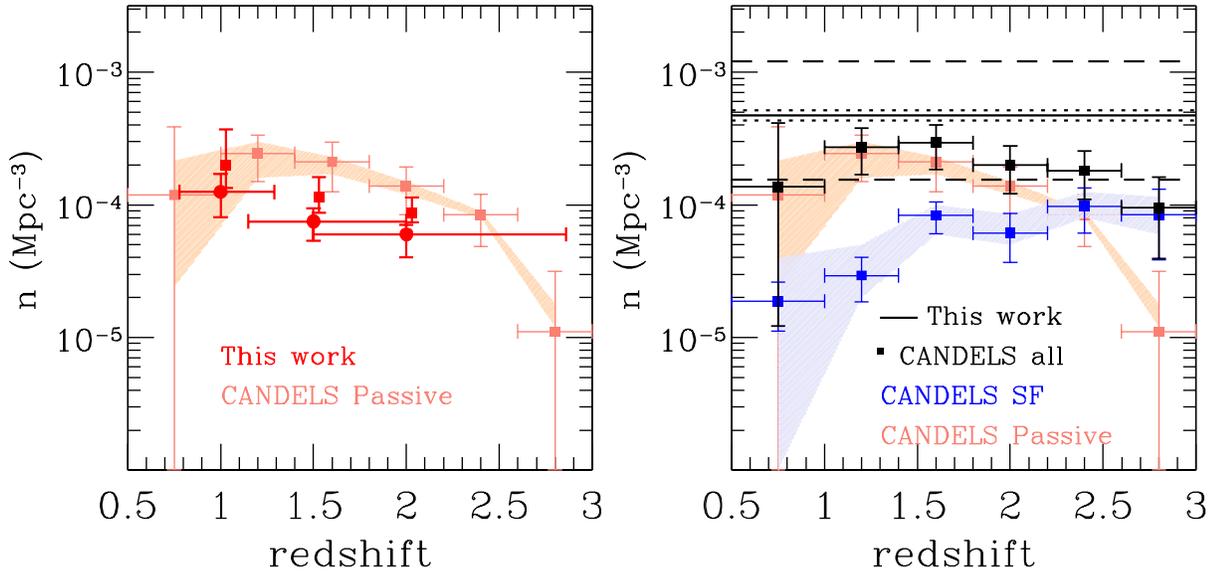}}
\vspace{-1.5cm}
\caption{Comparison of the number density of compact galaxies at high
and low redshift.  {\bf Left.} The number densities of {\it local}
(z=0.03-0.11)
PM2GC compact galaxies with LW ages compatible with being passive at
$z=1, 1.5$ and 2 are shown as red circles, and plotted at the
redshift corresponding to their LW age. Red rectangles show the values for
local compact galaxies adopting the
conversion to Maraston's masses at high-z and have been slightly shifted
in redshift for clarity.  Error bars on the Y axis
reproduce the values obtained for LW ages = LW age$_{z=1,1.5,2}$ $\pm
1$ Gyr. Error bars on the X axis are the $\pm 1$Gyr intervals around
each redshift.  Pink points are the number densities of CANDELS
compact passive galaxies observed at high redshift. The shaded
region encompasses the CANDELS number densities when their selection
thresholds in specific star formation and compactness limits are
modified by $\pm0.2$ dex, as given in B13.  {\bf
Right.}  The number density of {\it all} compact galaxies in the local
Universe, regardless of their LW age, is compared with the same
quantity observed by CANDELS at high redshift.  The PM2GC value is
shown by the horizontal solid black line, with Poissonian errors
(dotted lines) and upper and lower values obtained modifying the
compactness criterion by $\pm0.2$ dex (dashed lines).  The CANDELS
number density of all compact galaxies (black points) is obtained as
the sum of compact passive (pink) and compact star-forming (blue)
galaxies. The meaning of shaded areas is like in the left panel.}
\end{figure*}

The mild number density evolution of compact passive galaxies we
find in this paper is in good agreement with the results from Cassata
et al. (2011) who found that the number density of compact (at least
1$\sigma$ below the local mass-size relation) early-type passive
galaxies decreases only by a factor of two from $z\geq 1$ to $z=0$.

\subsection{The most extreme compact galaxies}

The compactness criterion and the mass range adopted in this paper 
are dictated by the B13 high-z sample (to which we aim to compare).
It would be desirable to assess the number density evolution
of galaxies of different masses and degree of compactness,
as soon as such detailed information will become available at high-z,
to investigate whether the evolution varies with galaxy mass and compactness.

As we have shown above, the B13 criterion corresponds to about 2$\sigma$
below the low-z median mass-size relation, which is similar
to what most previous works used to define compact galaxies.
This is the typical degree of compactness of high-z massive and
passive galaxies found in other studies as well, except for
the extremely compact galaxies in the sample at $z \geq 2$
of van Dokkum et al. (2008),
that are outliers in the stellar densities and radii distributions
(see P13 for a discussion).

We can attempt to compare our results to the work of
Cassata et al. (2011), who considered separately the evolution of
``ultra-compact'' {\it early-type} (morphologically selected)
passive galaxies, defined to have
sizes at least
0.4 dex below the local {\it early-type} mass-size relation and
representing about 20-30\% of their sample at $z \geq 1$.
If we consider PM2GC early-type galaxies
(ellipticals+S0s, Calvi et al. 2012) that are old in LW age ($z=1$) and lie at
least 0.4 dex below the PM2GC median mass-size relation for early-type
galaxies (Table~1 in P13), taking into account a 0.1 dex mass
evolution as before, we obtain a number density $0.4 \times 10^{-4}
\rm \, Mpc^{-3}$ and $0.9 \times 10^{-4} \rm \, Mpc^{-3}$ for Bruzual
\& Charlot and Maraston high-z models, respectively, again under the
assumption of no star formation activity at redshifts below 1.  For
reference, the ultracompact densities in Cassata et al. (2011) are
$\sim 2.2 \times 10^{-4} \rm \, Mpc^{-3}$ at $z=1-1.2$. 

This seems to suggest that the density evolution of the ultracompact 
population has been stronger than that of the whole compact population,
up to a factor between 2.5 and 5. Again, these are upper limits to the
evolution, assuming no star formation at later times.
This is based on a morphologically selected
sample, and it would be useful to repeat the comparison with 
large high-z samples of ultracompact galaxies of all morphological types
as was done in the rest of the paper.

It is also worth noticing that the most extreme compact galaxies at
high redshifts mostly lie at $z>2$ (Cassata et al. 2011, Cimatti et
al. 2012, P13 Fig.~4), and are likely observed soon after the end of
their star formation activity.  Under these conditions, the passive
evolution in mass we discussed in \S.2.1 can be very strong on a short
timescale.
A galaxy forming the bulk of its stars at $z=2.5$, between
$z=2-2.2$ and today will have lost 19-23\% of its $z=2-2.2$ observed
mass.  Already at $z=1$, such galaxy will have lost 13-17\% of its
$z=2-2.2$ mass.  The effect could be much larger for galaxies that
stopped forming stars just before being observed as passive at high
redshift, as suggested by the B13 results.
In general, passive mass evolution affects the comparison of
high- and low-z samples.
A stellar generation
evolving passively since high-z, by $z=0$ will have lost almost 50\%
of its initial mass. 

The exact amount of mass evolution depends on the
IMF assumed, on the epoch of major star formation and
detailed star formation history, therefore is highly uncertain.
The effect will become more and more prominent as observations
approach the epoch when star formation stopped.

We suggest that the extreme compactness of $z>2$ galaxies and the very
strong evolution between $z=2$ and $z=1$ might be at
least partially due to a fast passive evolution in {\it mass}
between $z=2$ and $z=1$.

\section{Conclusion}

We place an upper limit to the evolution in the number density of passively
evolving, compact galaxies from high redshift to the nearby Universe.
We conclude that at most half, and possibly an even smaller fraction,
of the high-z population has appreciably evolved in size. A more
precise estimate of the amount of size evolution is hindered mainly by
the uncertainty in high-z galaxy mass estimates, and by the
difficulty to match exactly progenitors and descendants on
the basis of luminosity-weighted ages. However, even under
conservative assumptions, our results do not demand a major
evolution in size for most of the compact high-z galaxies.
On the other hand, we find that the amount of evolution may be 
stronger (upper limits between 2 and 5) for the most extreme, 
ultracompact galaxies,
which represent a minority of the massive and passive population at high-z.
A detailed comparison as a function of galaxy mass and degree of
compactness will be feasible when the corresponding information
becomes available for high-z samples.






\begin{table*}
\caption{Number density of low-z compact galaxies}
\begin{center}
\begin{tabular}{lccc}
\hline
                & $LW age_{z=1}$ & $LW age_{z=1.5}$  & $LW age_{z=2}$ \\
         & \multicolumn{3}{c}{$10^{-4} \rm Mpc^{-3}$}\\
\hline
&&& \\
Compact (Bruzual\&Charlot)                & $1.3^{1.7}_{0.8}$  & $0.8^{0.9}_{0.5}$ & $0.6^{0.7}_{0.4}$ \\
($\rm log(M_{low-z}/r_e^{1.5}) \geq 10.2 \, M_{\odot} kpc^{-1.5}$) & & & \\ 
&&& \\
 Compact (Maraston)    & $2.0^{3.7}_{1.3}$  & $1.1^{1.6}_{0.9}$ & $0.8^{1.1}_{0.7}$ \\
($\rm log(M_{low-z}/r_e^{1.5}) \geq 10.05 \, M_{\odot} kpc^{-1.5}$) & & & \\ 
&&& \\
\hline 
\end{tabular}
\caption*{PM2GC number density values of compact galaxies at low
redshifts with LW ages $\geq$ than the lookback time to
the corresponding redshift (see text).  Values are given both adopting
the B13 compactness criterion that uses Bruzual \& Charlot models at
high-z, and converting the high-z masses to Maraston's models.  The
lower and upper values at each redshift were obtained varying the LW
age by $\pm 1$ Gyr.}
\end{center}
\end{table*}

\acknowledgments
We gratefully acknowledge useful discussions and the datapoints provided
by Guillermo Barro, and support from the WINGS team.
We thank Joe Liske, Simon Driver and the whole
MGC team for making their valuable dataset public and easily accessible.
We acknowledge financial support from the PRIN-MIUR 2009 and PRIN-INAF 2010.

\end{document}